\documentstyle[12pt,epsfig]{article}
\def\lsim{\raise0.3ex\hbox{$<$\kern-0.75em\raise-1.1ex\hbox{$\sim$}}}
\def\gsim{\raise0.3ex\hbox{$>$\kern-0.75em\raise-1.1ex\hbox{$\sim$}}}

\def\pom{{I\!\!P}}

\def\beq{\begin{equation}}
\def\eeq{\end{equation}}
\def\bea{\begin{eqnarray}}
\def\eea{\end{eqnarray}}
\def\bq{\begin{quote}}
\def\eq{\end{quote}}

\newcommand{\ssn}{\mbox{\boldmath $s$}}
\newcommand{\rr}{\mbox{\boldmath $r$}}
\newcommand{\rrn}{\mbox{$r$}}

\def\gappeq{\mathrel{\rlap {\raise.5ex\hbox{$>$}}
{\lower.5ex\hbox{$\sim$}}}}

\def\lappeq{\mathrel{\rlap{\raise.5ex\hbox{$<$}}
{\lower.5ex\hbox{$\sim$}}}}

\def\Toprel#1\over#2{\mathrel{\mathop{#2}\limits^{#1}}}

\newcommand{\rk}{\mbox{\boldmath $k$}}

\newcommand{\rkn}{\mbox{$k$}}
\def\pom{{I\!\!P}}

\begin{document}
\pagestyle{empty}
\begin{center}
{\bf PHENOMENOLOGY ON  THE QCD  DIPOLE PICTURE REVISITED}
\\
\vspace*{1cm}
 A.I.  Lengyel $^{1}$, M.V.T. Machado  $^{2,\,\,3}$\\
\vspace{0.3cm}
{$^{1}$ \rm Institute of Electron Physics, National
Academy of Sciences of Ukraine \\  Universitetska 21,
UA-88016 Uzhgorod, Ukraine  \\
$^{2}$ Instituto de F\'{\i}sica e Matem\'atica,  Universidade
Federal de Pelotas\\
Caixa Postal 354, CEP 96010-090, Pelotas, RS, Brazil\\
$^{3}$ \rm High Energy Physics Phenomenology Group, GFPAE,  IF-UFRGS \\
Caixa Postal 15051, CEP 91501-970, Porto Alegre, RS, Brazil}\\
\vspace*{1cm}
{\bf ABSTRACT}
\end{center}
  We perform an adjust to the most recent structure function data,  considering the QCD  dipole picture applied to  the $ep$ scattering. The structure function $F_2$ at small $x$ and intermediate $Q^2$ can be described by the model containing   an economical number of  free-parameters, which encodes the hard   Pomeron physics. The structure function $F_L$ and the gluon distribution are predicted without further adjustments. The data description is remarkably good, however giving an either low value for the fixed strong coupling constant. This  indicates that a resummed next-to-leading level analysis should be done, which would bring it to suitable values. 
\vspace*{1mm}
\noindent

\vspace*{1cm}
\noindent
\rule[.1in]{16.5cm}{.002in}

\vspace{-2cm}
\setcounter{page}{1}
\pagestyle{plain}

\vspace{1cm}

\section{Introduction}

Deep inelastic electron-proton scattering experiments at HERA have provided
measurements of the inclusive structure function $F_2(x,Q^2)$ in very
small values of the Bjorken variable $x$ ($10^{-2}$ down to $10^{-5}$). In
these processes the proton target is analyzed by a hard probe with virtuality
$Q^2=-q^2$, where $x \sim Q^2/2p.q$ and $p,\,q$ are the four-momenta of the
incoming proton and the virtual photon probe. In that kinematical region, the
gluon is the leading parton driving the small $x$ behavior of the deep
inelastic observables. The small $x$ region is described in a formal way using
the summation of gluon ladder diagrams, whose virtual contributions lead to the
gluon reggeization. Such ladders diagrams are associated
with the Pomeron, the leading  Reggeon  with the vacuum quantum numbers,  which
was introduced phenomenologically to describe the high energy behavior of
the total and elastic cross-sections of the hadronic reactions  and connected with the existence of large rapidity gaps in the produced final
state \cite{Predazzi}.

At the leading order (LO) all powers of $\alpha_s\ln(Q^2/\mu^2)$, with $\mu^2$
the factorization scale, are summed by the DGLAP evolution equations
\cite{DGLAP}, which take into account only the  strongly ordered parton
transverse momenta ($k_T$) ladders. At the present, the next-to-leading order (NLO)
contribution is also considered, including non-ordered $k_T$ contributions in a
covariant way. In the current HERA kinematical regime the DGLAP
approach is quite successful, although the theoretical expectation of
deviations due to the  parton saturation phenomena
could  be present \cite{saturation}. Following this point of view, recently the structure function measured at HERA has been studied considering an  analysis based on the first non-linear corrections to the DGLAP evolution \cite{Carlos}, showing for the gluon distribution the nonlinear effects play an increasingly role at $x\leq 10^{-3}$ and $Q^2\leq 10$ GeV$^2$. The procedure gives stable results for that  distribution at both low $x$ and $Q^2$, instead of negative or constant values as in the recent NLO DGLAP analysis (see \cite{Carlos} for details).

On the other hand, at very small $x$ the leading logarithms $\alpha_s\ln(1/x)$
are shown to be  important. In the leading logarithmic
approximation (LLA) the QCD Pomeron corresponds to the sum of ladder diagrams
with reggeized gluons along the chain, which are strongly ordered in momentum
fraction $x$. Such sum is described by the Balitzkij-Fadin-Kuraev-Lipatov
(BFKL) equation \cite{BFKL}. The corrections at next-to-leading level (NLLA)
are now known, leading to strong modifications in the LLA spectrum \cite{BFKLNLO}. At first glance, the results turned out conflicting and not useful for suitable phenomenological studies. Further, these features have been cured by canceling not desired singularities via a suitable resummation at all orders of the perturbative expansion, satisfying the renormalization group requirements \cite{resum1,resum2,resum3}. 
The main shortcoming in doing phenomenology with a BFKL evolution beyond the leading order is the complicated mismatch between QCD perturbative and non-perturbative inputs, as the corresponding factorization properties are more involved than the DGLAP evolution.

A very promising approach encoding all order  $\alpha_s\,\ln (1/x)$ resummation is provided by the QCD dipole picture. It was proven that such approach reproduces the  BFKL evolution \cite{dipole}. The main process is the onium-onium scattering, that is the reaction between two heavy quark-antiquark states, in such way that the process is perturbative since the onium radius provides the essential scale at which the strong coupling constant is evaluated. In the large $N_c$ limit, the heavy pair and the soft gluons are represented as a collection of color dipoles. The cross section is written as a convolution between the number of the dipoles in each onium state and the basic cross section for dipole-dipole scattering due to two-gluon exchange. The QCD dipole model can be applied to deep inelastic lepton-nucleon scattering, assuming that the virtual photon at high $Q^2$ can be described by an onium. On the other hand, the proton is approximately described by a collection of onia with an average onium radius to be determined from phenomenology. This model has produced a successful description of the old structure function data \cite{phenomenology}.  

The QCD dipole picture allows a systematic framework for testing the resummed next-to-leading order BFKL evolution kernels, producing predictions for the proton structure function. A method for doing this was recently proposed \cite{Peschanski}, where the resummation schemes are tested through the Mellin transformed $j$-moments of $F_2$. Moreover, it has been shown that a geometric scaling for the photon-proton cross section and the symmetry between low and high $Q^2$ regions are associated to the symmetry of the two-gluon dipole-dipole cross section \cite{munier1}. There, the proton is assumed to be a collection of independent dipoles at the time the interaction whose  sizes are distributed around $1/Q_s(x)$, with the latter being also the mean distance between the centers of the neighboring dipole. The quantity $Q_s^2 (x)=\Lambda^2 \,e^{\,\lambda \,\log\,(1/x)}$ is the saturation scale \cite{scale}. Further, it was shown that the {\it local} geometric scale can be testable experimentally \cite{munier2}. That is, the cross sections for exclusive processes only depend  on the ratio of scales $Q/Q_s(x,b)$, where now the saturation scale $Q_s^2(x,b)$ has a impact parameter ($b$) dependence.  

In this work we revisit the phenomenology using this successful approach. The analysis is organized as follows.  In the next section we shortly review the main formulae for the QCD dipole picture and give the phenomenological expressions for the structure functions and for the gluon distribution. In Sec. 3 we analyze the fitting procedure to the recent structure function data, determining the quality of the fit procedure and the range of applicability of the model. At the last section we drawn the main conclusions and summarize the results.

\section{The QCD color dipole picture and structure functions}

 Let us start by describing the main features and basic formulas  of the QCD dipole picture at high energies. The forward amplitude, ${\cal A}\,(s,t)$, for the onium-onium scaterring, integrated over impact parameter is written as,
\begin{eqnarray}
{\cal{A}}\,(s,t=0) & = &  -\, i\,\int d^2\rrn_{1}\,d^2\rrn_{2}\int dz_1\, dz_2 \,
\Phi^{(0)}\,(\rr_{1},\,z_1)\,\Phi^{(0)}\,(\rr_{2},\,z_2) \nonumber \\
& & \times \int \,\frac{d^2s_1}{\ssn_1}\,\frac{d^2s_2}{\ssn_2}\,n\,(Y/2,\,\rr_1,\,\ssn_1)\, n\,(Y/2,\,\rr_2,\,\ssn_2)\, \sigma_{\mathrm{dd}}\,(\ssn_1,\, \ssn_2)\,,
\label{ampli1}
\end{eqnarray}
where $\Phi^{(0)}(\rr_{i},z_i)$ is the squared wave function of 
the quark-antiquark part of the  onium wave function,
$\rr_{i}$ being the transverse size of the quark-antiquark pair 
and $z_i$ the longitudinal momentum fraction of 
the antiquark. The rapidity is given by $Y/2 \simeq \ln (x_0/x)$, where $x_0$ is a phenomenological constant and $x$ is the Bjorken variable labelling the softer end of the produced dipole. The dipole density is labeled as $n\,(Y/2,\rr_i,\ssn_i)$, which is discussed below. The quantity   $\sigma_{\mathrm{dd}}$  is the elementary 
dipole-dipole cross-section and reads as,
\begin{eqnarray}
 \sigma_{\mathrm{dd}}\,(\ssn_1,\ssn_2) = 2\,\pi\,\alpha_s^2\,\, [\mathrm{min} \,(\ssn_1,\ssn_2)]^2\,\left[1 - \ln \left( \frac{\mathrm{max}\,(\ssn_1,\ssn_2)}{\mathrm{min}\,(\ssn_1,\ssn_2) } \right) \right]\,.
\end{eqnarray}

In the large $N_c$ limit and in the leading-logarithmic-approximation the 
radiative corrections are generated by emission of gluons with strongly 
ordered longitudinal momenta fractions $z_i >> z_{i+1}$. The onium wave 
function with $n$ soft gluons  can be calculated using perturbative QCD. In 
the Coulomb gauge the soft radiation can be viewed as a cascade of color 
dipoles emitted  from the initial  dipole, since each gluon 
acts like a quark-antiquark pair. The dipole density $n(Y,\rr,\ssn)$ was defined  \cite{dipole} in such way,
\begin{eqnarray}
N\,(Y,\ssn) = \int dz \int d^2 \rrn  \,\Phi^{(0)}(\rr,z)\,\, n(Y,\rr,\ssn)
\end{eqnarray}
is the number of dipoles of transverse size $\ssn$ with the smallest 
light-cone momentum in the pair   greater than or equal to $e^{-Y}p_+$, 
where $p_+$ is the light-cone momentum of the onium. The whole dipole cascade can be constructed from a 
repeated action of a kernel ${\cal{K}}$ on the initial density 
$n_0\,(\rrn,\ssn)$ through the dipole 
evolution equation, 
\begin{eqnarray}
n\,(Y,\rr,\ssn_1) = n_0\,(\rr,\ssn_1) +
\int_0^Y dy \int_0^{\infty} \, d\ssn_2 \,\, {\cal{K}}\,(\ssn_1,\ssn_2) \,n\,(y,\rr,\ssn_2)\,.
\label{ker_bfkl}
\end{eqnarray}

The evolution kernel $\cal{K}$ is calculated in perturbative QCD. For 
fixed $\alpha_s$ and in the limit of large $N_c$ the kernel has 
the same spectrum as the BFKL kernel \cite{dipole}. Therefore, the two approaches lead to
the same phenomenological results for inclusive observables.
The solution of Eq. (\ref{ker_bfkl}) is given by \cite{dipole},
\begin{eqnarray}
n\,(Y,\rr,\ssn) = \frac{1}{2} \,\frac{\rr}{\ssn} \frac{\exp \left( \omega_{\pom}\,Y\right)}{\sqrt{7\, \pi\, \bar{\alpha}_s \, \zeta (3)\,Y}}\,\,
\exp\,\left(-\frac{ \ln^2 \, (\rr/\ssn)}{28\, \bar{\alpha}_s \, \zeta (3)\,Y}\right)\,,
\label{nsol}
\end{eqnarray}
where $\omega_{\pom}=4\,\bar{\alpha}_s \ln 2$, with $\bar{\alpha}_s=\alpha_s N_c/\pi$,   and $\alpha_{\pom}=1 + \omega_{\pom}$ is the Pomeron intercept. Therefore,  the cross section grows rapidly with the energy since the number of dipoles in the light cone wave function grows rapidly on energy. The result in Eq. (\ref{nsol}) has been obtained considering a  process having  only one scale, that is the onium radius. In the electron-proton deep inelastic scattering  two scales are present and this result should be  affected by non-perturbative  contributions. The hard scale is given by the photon virtuality and the soft one is associated to the proton typical size.

Now, let us review the obtention of the  proton structure functions in the QCD dipole picture.  In order to do so, an assumption is made that the proton can be approximately described by onium configurations. The basic assumption is given by, 
\begin{eqnarray}
\sigma^{\gamma\, p}_{L,T}\,(x,Q^2) = \sigma^{\gamma \,\,onium}_{L,T}\,(x,Q^2,\rk^2; \mu_F^2)\, \tilde{n}\,(\rk^2; \mu_F^2)\,,
\label{eqmaster}
\end{eqnarray}
where $\tilde{n}\,(k^2;\,\mu_F^2)$ is the probability of finding an onium in the proton, with  $\mu_F^2$ being a unknown perturbative scale characterizing the average onium size, as will be seen later on. We follow Refs. \cite{phenomenology}  to write down the virtual photon-onium cross section, where the $k_T$-factorization approach \cite{kt_fact} was considered within the framework of  the QCD dipole model. It reads as, 
\begin{eqnarray}
\sigma^{\gamma \,\,onium}\,(x,Q^2) =  \int d^2 \rrn \, dz \,\Phi^{(0)}\,(\rr,\,z)\,\, \sigma^{\gamma-dipole}\,(x,Q^2,\rr)\,,
\label{sigma1}
\end{eqnarray}
where $\Phi^{(0)}\,(\rr,\,z)$ was already defined before. The virtual photon-dipole cross section is written as,
\begin{eqnarray}
Q^2\,\sigma^{\gamma-dipole}\,(x,Q^2,\rr) = \int^{Q^2}\, d^2\rkn \,\int_0^1 \frac{dz}{z} \,\sigma_{\gamma\,g}\left(\,\frac{x}{z},\, \frac{\rk^2}{Q^2}\right)\,{\cal{F}}_{onium}\,(z,\rk, \rr)\,\,\,.
\label{sigma2}
\end{eqnarray}
where  $\sigma_{\gamma \, g}/Q^2$ is the Born cross section of an off-shell gluon of transverse momentum $\rk$ in the the subprocess $\gamma^* g \rightarrow q \overline{q}$ and  ${\cal{F}}_{onium}$  is the unintegrated gluon distribution function of an onium state of size $\rrn$, containing the hard Pomeron physics. The relation between this quantity  and  the  dipole density is written as, 
\begin{eqnarray}
\rk^2 \,{\cal{F}}_{onium}\,(z,\rk, \rr) = \int \,\frac{d\ssn^2}{\ssn^2}\, \int_0^1\, \frac{dz^{\prime}}{z^{\prime}}\, n\,(z^{\prime} ,\rr, \ssn) \, \,\sigma^{\gamma-dipole}\,\left(\frac{z}{z^{\prime}}\,, \,(\ssn\,\rk)^2 \right)\,,
\label{sigma3}
\end{eqnarray}
where $n\,(z^{\prime}, \rr, \ssn)$ is the density of dipoles of transverse size $\ssn$ with the smallest light-cone momentum in the pair equal to $z^{\prime} p_+$ in a dipole of transverse size $\rr$, of total momentum $p_+$.  This is 
given by the solution in Eq. (\ref{nsol}).

The next step is to apply a Mellin transform in the $x$-space and further a second Mellin transform in the $\rkn^2$-space, where the conjugated variables are $\omega$ and $\gamma$, respectively. After that, one obtains \cite{phenomenology},  
\begin{eqnarray}
Q^2\,\sigma^{\gamma-dipole}\,(x,Q^2,\rr) = \int\, \frac{d\,\omega}{2\,i\,\pi}\, \int\, \frac{d\,\gamma}{2\,i\,\pi}\,\ \sigma_{\gamma\,g}\,(\omega, 1-\gamma)\, {\cal F}_{onium}\,(\omega, \,\gamma) \, (Q^2\,\rr^2)^{\gamma}\, x^{-\,\omega}\,.
\label{sigma4}
\end{eqnarray}

The unintegrated gluon distribution for an onium of size $\rrn$ can be obtained from perturbative QCD and it is written as \cite{dipole},
\begin{eqnarray}
 {\cal F}_{onium}\,(\omega, \,\gamma) = \bar{\alpha}_s \, \frac{1}{\gamma}\,\frac{\nu \,(\gamma)}{\omega - \bar{\alpha}_s\, \chi\,(\gamma)}\,\,,
\label{fonium}
\end{eqnarray}
where in the leading-order  BFKL approach, $\chi_{\mathrm{L0}}(\gamma)=2\,\psi(1)-\psi\,(\gamma)-\psi\,(1-\gamma)$ and $\bar{\alpha}_s\,\chi\,(\gamma)$ is the eigenvalue of the LO BFKL kernel. The Mellin transform of the gluon-dipole coupling, $\nu \, (\gamma)$, can be obtained from $k_T$-factorization approach as,
\begin{eqnarray}
\nu \, (\gamma ) = \frac{2^{-\,2\gamma-1}}{\gamma}\,\frac {\Gamma\, (1-\gamma)}{\Gamma\, (1-\gamma)}\,.
\end{eqnarray}

The unintegrated gluon distribution, Eq. (\ref{fonium}), has a pole at $\omega_{\pom}= \bar{\alpha}_s\,\chi\,(\gamma)$ and then the integration over $\omega$ can be done analytically. Furthermore, the cross section $\sigma_{\gamma \,g}$ can be written in function of the LO photon impact factors $h_{L,\,T}(\gamma)$, corresponding to the perturbative coupling to the photon. Namely, $\sigma_{\gamma \,g} \equiv 4\,\pi^2 \, \alpha_{em}\,e_f^2 \, h_{L,\,T}\,(\gamma)$, with $e_f^2$ being the total charge of the quarks flavor contributing to the reaction  and where,
\begin{eqnarray}
\pmatrix{h_T \cr 
h_L}=\frac{\bar{\alpha}_s}{9\,\gamma}\frac{\left[\, \Gamma\,(1-\gamma)\,\Gamma\,(1+\gamma)\,\right]^3}{\Gamma
\,(2-2\gamma)\,\Gamma\,(2+2\gamma)}\frac{1}{1-\frac{2}{3}\,\gamma}\pmatrix{(1+\gamma)\,(1-
\frac{\gamma}{2}) \cr \gamma\,(1-\gamma)}\,,
\label{wfunctions}
\end{eqnarray}

Putting  Eqs. (\ref{wfunctions}) and (\ref{fonium}) in Eq. (\ref{sigma4}) and solving analytically on $\omega$ for the pole on $\omega=\omega_{\pom}$, one obtains,
\begin{eqnarray}
Q^2\,\sigma^{\gamma-dipole}\,(x,Q^2,\rr) = 8\, \pi^2\, \alpha_{em}\,e_f^2\,\bar{\alpha}_s \,\int\, \frac{d \gamma}{2\, \pi i} \, \, h_{L,T}\,(\gamma)\, \frac{\nu\,(\gamma)}{\gamma}\,\left(\rr^2\,Q^2 \right)^{\gamma}\, \exp \left[ \bar{\alpha}_s \, \chi \,(\gamma)\, \ln \frac{x_0}{x} \right]\,. 
\label{lastsig}
\end{eqnarray}

The virtual photon-onium cross section is obtained replacing  Eq. (\ref{lastsig}) in Eq. (\ref{sigma1}). The result depends on the   squared wave function $\Phi^{(0)}$  of the onium state, which cannot be computed perturbatively. This is  solved by  eliminating that dependence through an averaging over the wave function  of transverse size \cite{phenomenology},
\begin{eqnarray}
(\,\mu_F^2\,)^{-\,\gamma} = \langle \rr^2 \rangle \equiv  \int\, dz \,d^2\rrn \,\,(\rr^2)^{\gamma}\, \Phi^{(0)}\, (\rr,\,z)\,\,,
\end{eqnarray}
where $\mu_F^2$ is a scale which is assumed to be perturbative. With this result, the photon-onium cross section, Eq. (\ref{sigma1}), takes the following form,
\begin{eqnarray}
\sigma^{\gamma\,\, onium}\,(x,Q^2) = \frac{4\, \pi^2\, \alpha_{em}}{Q^2}\,2\,\bar{\alpha}_s \,\int\, \frac{d \gamma}{2\, \pi i} \, \, h_{L,T}\,(\gamma)\, \frac{\nu\,(\gamma)}{\gamma}\,\left(\,\frac{Q^2}{\mu_F^2} \,\right)^{\gamma}\, \exp \left[ \bar{\alpha}_s \, \chi \,(\gamma)\, \ln \frac{x_0}{x} \right]\,. 
\label{gammaonium}
\end{eqnarray}

In order to compute the photon-proton cross section one needs to know  the probability of finding an onium in it, $\tilde{n}\,(\gamma; \, \mu_F^2)$. Relying on renormalization group  properties, namely the overall result should be independent of the scale perturbative $\mu_F^2$, an suitable ansatz is given by \cite{phenomenology},
\begin{eqnarray}
\tilde{n}\,(\gamma;\,\mu_F^2) = n_{\mathrm{eff}}\,(\gamma)\,\left(\, \frac{\mu_F^2}{Q_0^2}\, \right)^{\gamma}\,\,,
\label{scale}
\end{eqnarray} 
where $Q_0^2$ is interpreted as a characteristic non-perturbative scale of the hadron-onium scattering. The quantity  $n_{\mathrm{eff}}$ means the average number of primary dipoles in the proton and $r_0 \equiv 2/Q_0$ their average transverse diameter (see second reference on \cite{phenomenology}). Placing the ansatz Eq. (\ref{scale}) in Eq. (\ref{eqmaster}) and using the result on Eq. (\ref{gammaonium}), the proton structure functions can be calculated. The integration over $\gamma$ can be performed using the steepest descent method, with the saddle point given by $\chi^{\prime}\,(\gamma_s)=-\ln (Q^2/Q_0^2)/\bar{\alpha}_s\,\ln (x_0/x)$. Using the expansion of the BFKL kernel near $\gamma = \frac{1}{2}$, one obtains,
\begin{eqnarray}
\gamma_s = \frac{1}{2} \left[ 1 - \kappa\,(x)\,\ln \left( \frac{Q}{Q_0} \right)\right]\,\,,
\end{eqnarray}
where $\kappa\,(x)=[\bar{\alpha}_s\, 7\, \zeta (3)\, \ln \frac{x_0}{x}]^{-1}$ is the diffusion coefficient at rapidity $Y=\ln \,(x_0/x)$. Putting all together, the structure functions are written in the simple form \cite{phenomenology},
 \begin{eqnarray}
F_{T,\,L}\,(x,Q^2) = H_{T,\,L}\, \frac{\bar{\alpha}_s\,\pi^3 \,e_f^2\,n_{\mathrm{eff}}}{96}\,\left(\, \frac{x_0}{x} \, \right)^{\omega_{\pom}}\, \frac{Q}{Q_0}\, \sqrt{2\,\kappa\,(x)/\pi}\,\exp \left[ -\frac{\kappa\,(x)}{2}\,\ln^2 \frac{Q}{Q_0} \right]\,\,,
\label{sfs}
\end{eqnarray}
where $H_T=9/2$ and $H_L=1$, defining $n_{\mathrm{eff}}\equiv n_{\mathrm{eff}}\,(\gamma_s)$ and $r_0\equiv r_0\,(\gamma_s)$. In the nest section we consider the equation above to determine the parameters of the model by fitting the recent data on the proton structure function $F_2=F_T + F_L$. The conditions to obtain the structure functions from the saddle-point method fix the conditions of its safe applicability. Hence, the following bound should be covered,
\begin{eqnarray}
\kappa \,(x)\,\ln \left( \frac{Q}{Q_0} \right)\simeq \frac{\ln \left(\frac{Q}{Q_0}\right)}{\ln \left(\frac{x_0}{x}\right)}\ll 1\,,
\end{eqnarray}  
which is realized for the region of moderate $Q/Q_0$ when compared to the range on $x_0/x$. It is worth to mention that at higher $Q^2$, the DLLA becomes valid, with the pole now at $1/\gamma$ in the kernel $\chi \,(\gamma)$.

\section{Fitting results and discussions}

In this section we will perform a fitting procedure using the recent HERA data  on the proton structure function \cite{H1rec,ZEUSrec} using the QCD dipole picture given by Eq. (\ref{sfs}). Having determined the parameters of the model from data, the longitudinal structure function and the proton gluon distribution are calculated without further adjustments. We made an simple attempt to describe phenomenologically the charm structure function, although the full expressions for it to be known in the literature (last reference in \cite{phenomenology}).

\begin{table}[t]
\begin{center}
\begin{tabular}{||c|c|c|c|c||}
\hline
\hline
& $\mathrm{PARAMETER}$  &   I       &  II     & III   \\
\hline
& ${\cal N}_p$ & 0.0625  & 0.0767 & 0.0985  \\
QCD dipole &  $Q_0$  & 0.464  & 0.544  & 0.587  \\
& $x_0$ & 1.90 & 1.0 (fixed) & 1.0 (fixed) \\
& $1+\omega_{\pom}$ & 1.25  & 1.26 & 1.23 \\
\hline 
& $A_R$ & 1.10 & 1.23  & - \\
& $a_R$ & -0.198  & -1.01  & - \\
Non-singlet & $d$ & -1.16 &  -0.416 & -  \\
& $b$ & -0.482 & -0.007 & -   \\
\hline
\hline
$\chi^2/\mathrm{d.o.f.}$ & & 1.26  &  1.25 & 1.02 \\
\hline
\hline
\end{tabular}
\end{center}
\caption{Parameters of the fit for the H1 data analysis.}
\end{table}

For sake of simplicity, we have defined the overall normalization for the proton structure function, ${\cal N}_p = (H_T + H_L)\,\bar{\alpha}_s\pi^3 e_f^2 \,n_{\mathrm{eff}}/96$.  Also, for some fitting procedures we will include a non-singlet contribution and threshold (large $x$) effects. Namely, the hard Pomeron expression, Eq. (\ref{sfs}), is multiplied by the factor $(1-x)^{n_s}$, while the non-Pomeron contribution is multiplied by $(1-x)^{n_{ns}}$. The expressions for the non-singlet piece and $n_{s,\,ns}$  read as,
\begin{eqnarray}
 F_2^{\mathrm{ns}}(x,Q^2) & = & A_R\, x^{1-\alpha_{R}}\,\left(\frac{Q^2}{Q^2+a_R}\right)^{\alpha_{R}}\,(1-x)^{n_{ns}({Q^2})}\, \label{softreg} \\
 n_{ns}(Q^2) & = & \frac{3}{2}\,\left(1+\frac{Q^2}{Q^2+d}\right)\,, \hspace{1cm} {n_s}(Q^{2})=\frac{7}{2} \,
\left(1+\frac{Q^{2}}{Q^{2}+b}\right)\,,
\label{nsinglet}
 \end{eqnarray}
where $\alpha_R=0.4$ is the Reggeon intercept, considered fixed in this analysis. Other values can be considered, however the  $\chi^2$ quality of the fit is worst.

Here, we have considered the following  fitting procedures:

\begin{itemize}

\item  (I) Adjusting data in the whole range $1.5 \leq Q^2 \leq 150$ GeV$^2$ and all $x$, considering the QCD dipole, Eq. (\ref{sfs}),  plus the non-singlet contribution, Eq. (\ref{softreg}). The motivation is to include the large $x$ data at high $Q^2$, where the fixed target data (E665 and NMC points \cite{E665,NMC})  were also added. The results for H1 and ZEUS data sets are presented in Tables (1) and (2). The number of points is equal to 230 (H1) and 259 (ZEUS), respectively. The total number of parameters is 7;

\item  (II)  Adjusting data in the  range $1.5 \leq Q^2 \leq 60$ GeV$^2$ and all $x$, considering the QCD dipole plus the non-singlet contribution. We have fixed the parameter $x_0=1$ in this case, reducing the final number of parameters. The results for H1 and ZEUS data sets are presented in Tables (1) and (2);

\item  (III) Adjusting data in the whole range $1.5 \leq Q^2 \leq 150$ GeV$^2$ and  restricting to small region  $x\leq 10^{-2}$, considering only the QCD dipole contribution. We have considered the fixed values $x_0=1$ and $n_s=7$.

\end{itemize}
\begin{table}[t]
\begin{center}
\begin{tabular}{||c|c|c|c|c||}
\hline
\hline
& $\mathrm{PARAMETER}$  &   I       &  II     & III   \\
\hline
& ${\cal N}_p$ & 0.0581  & 0.0638 & 0.0977  \\
QCD dipole &  $Q_0$  & 0.475  & 0.517  & 0.571  \\
& $x_0$ & 1.30  & 1.0 (fixed) & 1.0 (fixed) \\
& $1+\omega_{\pom}$ & 1.28  & 1.29 & 1.24 \\
\hline 
& $A_R$ & 1.42 & 1.43  & - \\
& $a_R$ & 0.383  & -0.273  & - \\
Non-singlet & $d$ & -0.842 &  -0.808 & -  \\
& $b$ & 1.36 & -0.677 & -   \\
\hline
\hline
$\chi^2/\mathrm{d.o.f.}$ & & 1.27  &  1.25 & 1.08 \\
\hline
\hline
\end{tabular}
\end{center}
\caption{Parameters of the fit for the ZEUS data analysis.}
\end{table}

Lets start discussing the procedure (I). The main difference between the two data sets are the values for the parameter $x_0$, being larger for ZEUS than for H1. On the other hand, the effective intercept for the hard QCD contributions is quite the same, that is $1+\omega_{\pom}=1.25\,(8)$, with the quality of fit better for ZEUS data set. It is worth to mention that the low values obtained for the parameters $b$ and $d$ suggests that they can be fixed in the usual way, $n_s=7$ and $n_{ns}=3$, in the range considered here. This procedure would reduce strongly the number of free parameters. From the fit, it is also clear the non-singlet piece is $Q^2$ independent, since $a_R$ is quite small.

For the procedure (II), the picture is similar to the previous one, with the effective intercept $1+\omega_{\pom}=1.26\,(9)$  coming out different  for the two data set. The quality of fit is also slightly equal in the two cases (see Tables). The conclusion about $n_{s,\,ns}$ is the same as the procedure (I).

Finally, in the procedure (III) we have considered only the small $x\leq 10^{-2}$ data (102 data points for H1 and 107 for ZEUS). In this case,  $x_0=1$ and $n_s=7$ are taken as fixed, reducing the number of parameters equal to 3. Indeed, their values are in large extent similar with a better quality of fit for the H1 data set. This procedure is similar to the previous analysis on Refs. \cite{phenomenology}, with an even  lower effective power ($1+\omega_{\pom}\simeq 1.282$ in \cite{phenomenology}). The low value for the fixed coupling constant $\alpha_s\simeq 0.1$ reveals the well known necessity of sizeable higher order corrections to the approach. Accurate analysis in this lines, considering resummed NLO BFKL kernels, has been proposed recently \cite{Peschanski} producing a reasonable $\alpha_s\sim 0.2$ for the typical $Q^2$ range considered in phenomenology for structure functions.

\begin{figure}[t]
\centerline{\epsfig{file=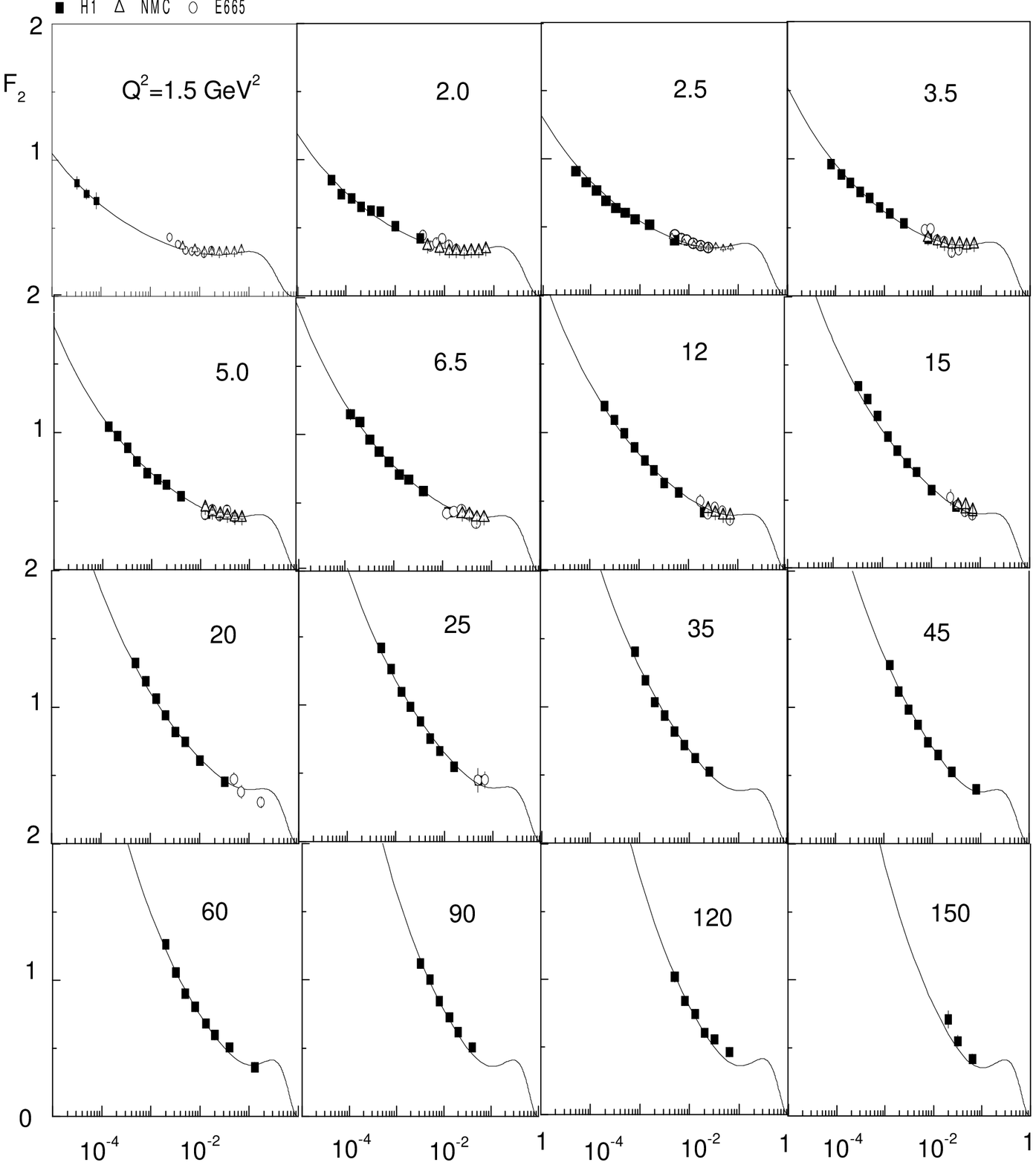,width=13cm,height=16cm}}
\caption{The inclusive structure function $F_2$ for the H1 data set \cite{H1rec} and fixed target data points \cite{E665,NMC}. The curve corresponds to procedure (II).}
\label{fig1}
\end{figure}

\begin{figure}[t]
\centerline{\epsfig{file=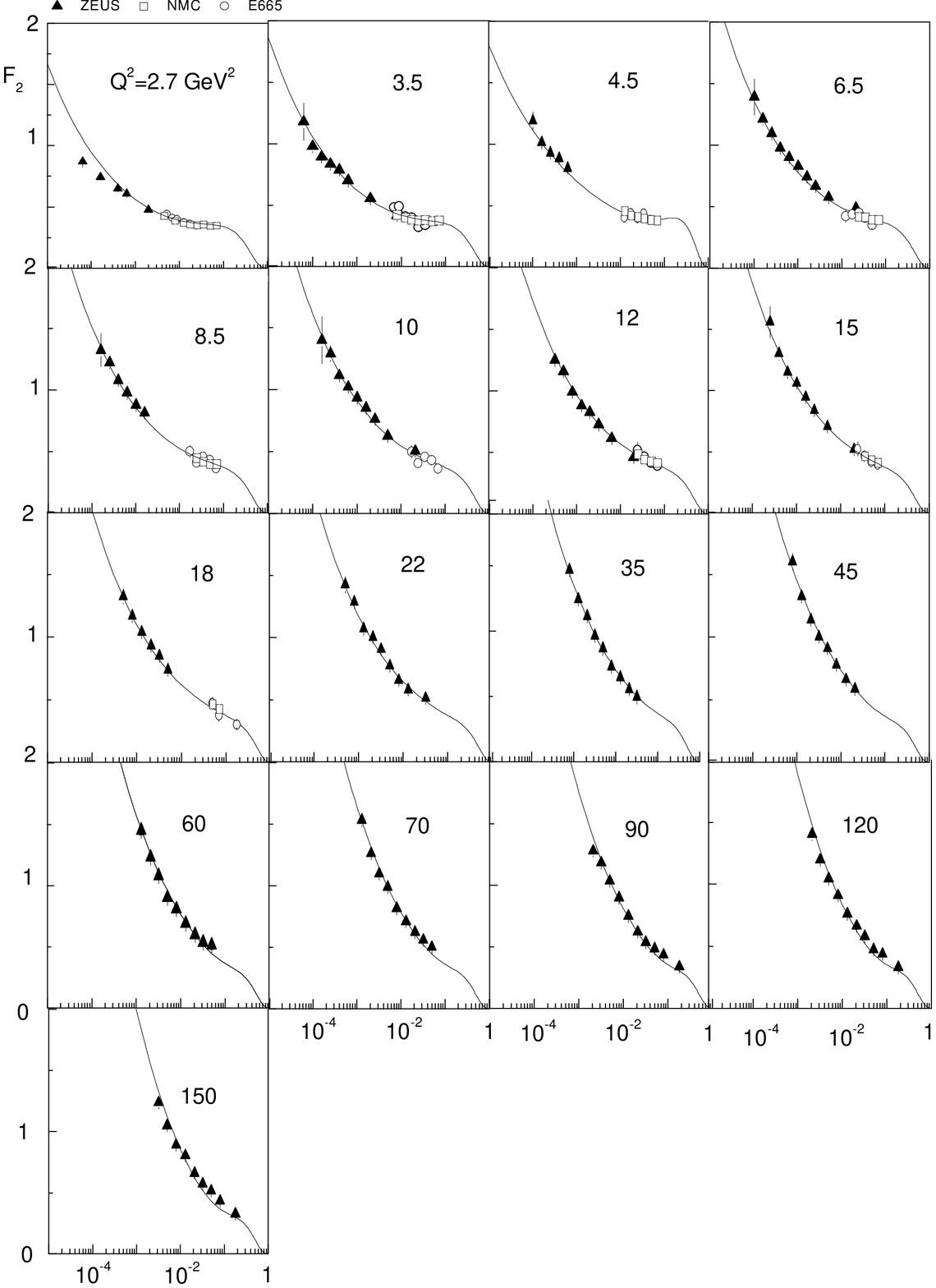,width=13cm,height=16cm}}
\caption{The inclusive structure function $F_2$ for the ZEUS data set \cite{ZEUSrec} and fixed target data points \cite{E665,NMC}. The curve corresponds to procedure (II).}
\label{fig2}
\end{figure}

\begin{figure}[t]
\centerline{\epsfig{file=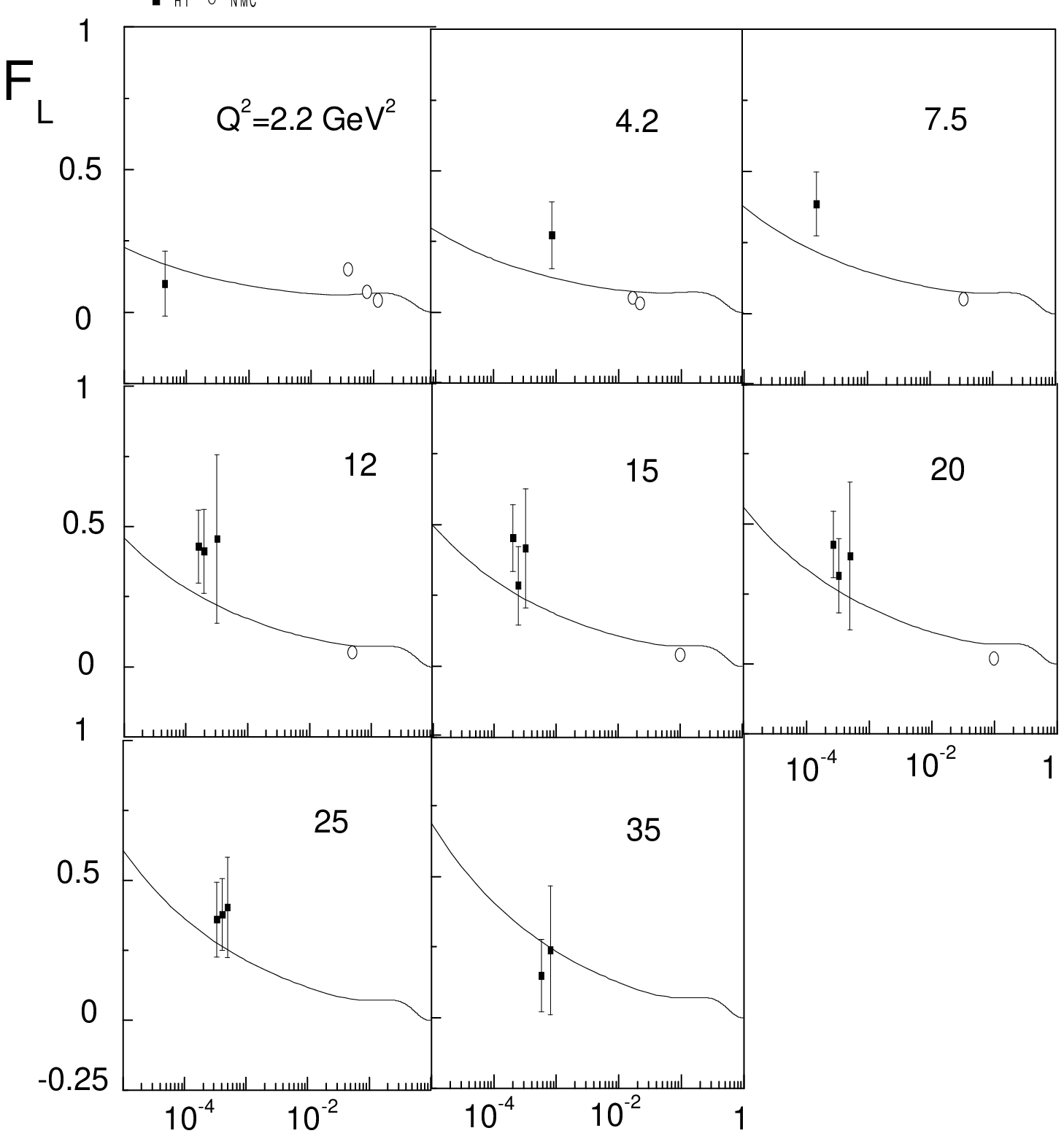,width=13cm,height=19cm}}
\caption{The longitudinal structure function $F_L$ \cite{H1rec}, where the parameters of procedure II (H1 data set) were taken into account and considering $F_L=(2/11)\,F_2$.}
 \label{fig4}
\end{figure}

In order to illustrate the analysis, in Figs. \ref{fig1} and  \ref{fig2} we present the fit result (procedure II) for the inclusive structure function for  H1 \cite{H1rec} and ZEUS \cite{ZEUSrec} data set, including the fixed target points  \cite{E665,NMC} and extrapolated up to $Q^2=150$ GeV$^2$. It is worth to mention that the new data are more precise than the previous analysis and the same quality of fit is obtained in a small range on $Q^2$, in contrast with the studies in \cite{phenomenology}. However, a rough data description can be obtained up to higher values. Here, our aim is discuss the model as the quality of fit. We have checked using the new H1 data that a $\chi^2/{\mathrm{dof}}=1.4$ is obtained in the range $1.5\leq Q^2\leq 150$ for all $x$. This in some extent shows the range on virtuality where a LO BFKL approach is suitable. That is, low $x$ and not so high $Q^2$. In virtualities of hundreds of GeV, the physics should be described by the DLA limit of the BFKL approach and which is a common limit also in the DGLAP evolution \cite{Victor}. An interesting further study is to realize a best fit test to the DLA limit, determining its range of applicability.  After this discussion, having determined the parameters of the model from data on $F_2$, it is possible determine without further adjustments the structure function $F_L$ and the gluon distribution $x\,G(x,Q^2)$. From Eq. (\ref{sfs}) and the definition of the overall normalization, we can see that $F_L=(2/11)\,F_2$, since $H_T+H_L=11/2$ and $H_L=1$. In Fig. (\ref{fig4}) this result is shown using parameters from procedure II from the  H1 data set against the recent $F_L$ data \cite{H1rec}. The results are in good agreement with the recent data, obtained without further adjustments.

\begin{figure}[t]
\centerline{\epsfig{file=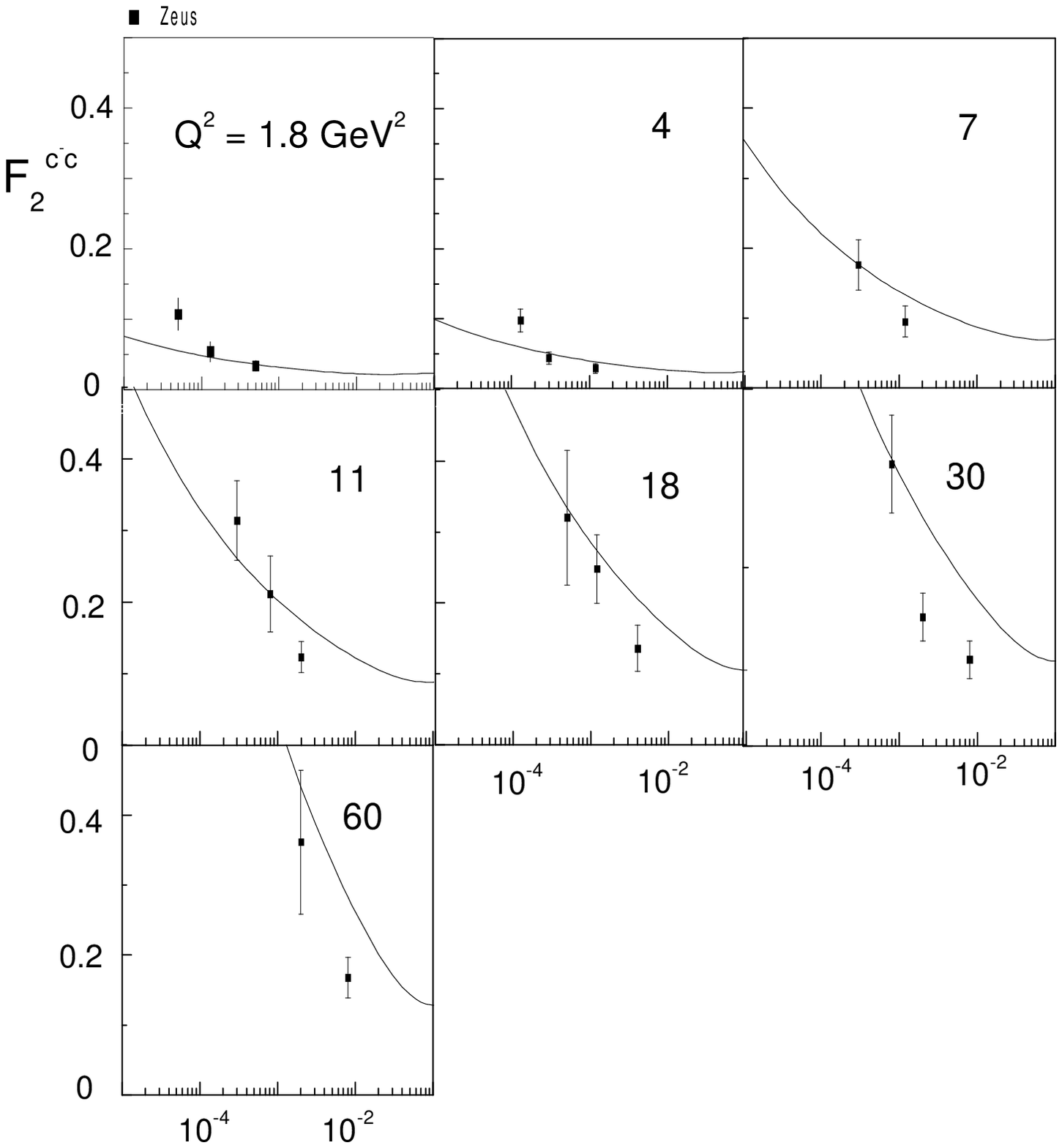,width=14cm,height=19cm}}
\caption{The charm structure function $F_2^{c\bar{c}}$, where the parameters of procedure II (ZEUS data set) were taken into account.}
 \label{fig6}
\end{figure}

We made an attempt to describe phenomenologically the charm structure function, despite explicit expressions for it to be available. One suppose the following relation between the $F_2$ structure function and $F_2^{c\bar{c}}$,
\begin{eqnarray}
F_2^{c\bar{c}}\, (x,Q^2)= \frac{2}{5}\,\left( 1+ \frac{4\,m_c^2}{Q^2}\right)^{-1}\,F_2\,(x,Q^2)\,\,, 
\end{eqnarray}
where we have used $m_c=1.5$ GeV and the factor $2/5$ corresponds to $e_c^2/(\sum \,e_f^2+e_c^2)$ . The result is shown in Fig. (\ref{fig6}), where the parameters from procedure II (ZEUS data set) were taken into account. The simple ansatz is reasonable only at low $Q^2$, overestimating the data on higher virtualities.  It should be stressed that this function can be calculated using the complete expressions for the photon impact factors, $h_{T,L}\,(\gamma; m_f)$, which consider also the dependence on the quark masses, $m_f$ (see last reference in \cite{phenomenology}).

As a final analysis, in the QCD dipole approach the gluon distribution function can be calculated in a straightforward way from $F_2$. The result is also independent of the overall normalization, which contains part of the non-perturbative inputs of the model. The gluon function is given by,
\begin{eqnarray}
x\,G(x,Q^2) & = &  \left[h_T\,(\gamma=\gamma_s)+ h_L\,(\gamma=\gamma_s) \right]^{-1}\,F_2(x,Q^2)\,,\nonumber\\
& = & \frac{9\,\gamma_s}{\bar{\alpha}_s}\,\frac{1- \frac{2}{3}\,\gamma_s}{1+ \frac{3}{2}\,\gamma_s - \frac{3}{2}\,\gamma_s^2}\,\, \frac{\Gamma\,(2-2\,\gamma_s)\,\Gamma\,(2+2\,\gamma_s)}{\left[\, \Gamma\,(1-\gamma_s)\,\Gamma\,(1+\gamma_s)    \,  \right]^3}\,F_2(x,Q^2)\,.
\label{xgxq2}
\end{eqnarray}

\begin{figure}[t]
\centerline{\epsfig{file=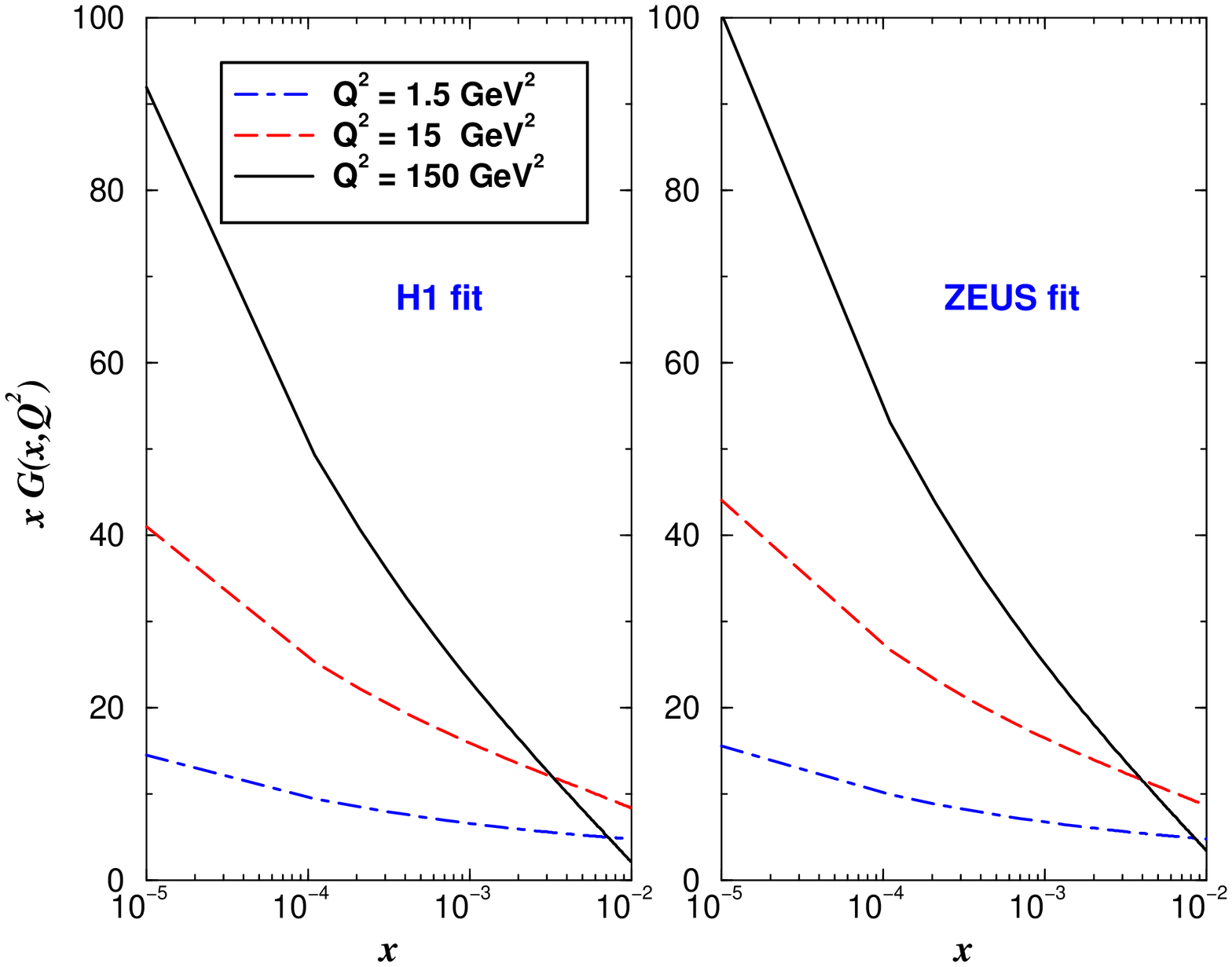,width=14cm,height=13cm}}
\caption{The gluon distribution function $x\,G(x,Q^2)$,  where the parameters of procedure III  were taken into account. In the plot on the left, parameters from H1 data set fit and on the right the ZEUS data set fit.}
 \label{fig7}
\end{figure}

\begin{figure}[t]
\centerline{\epsfig{file=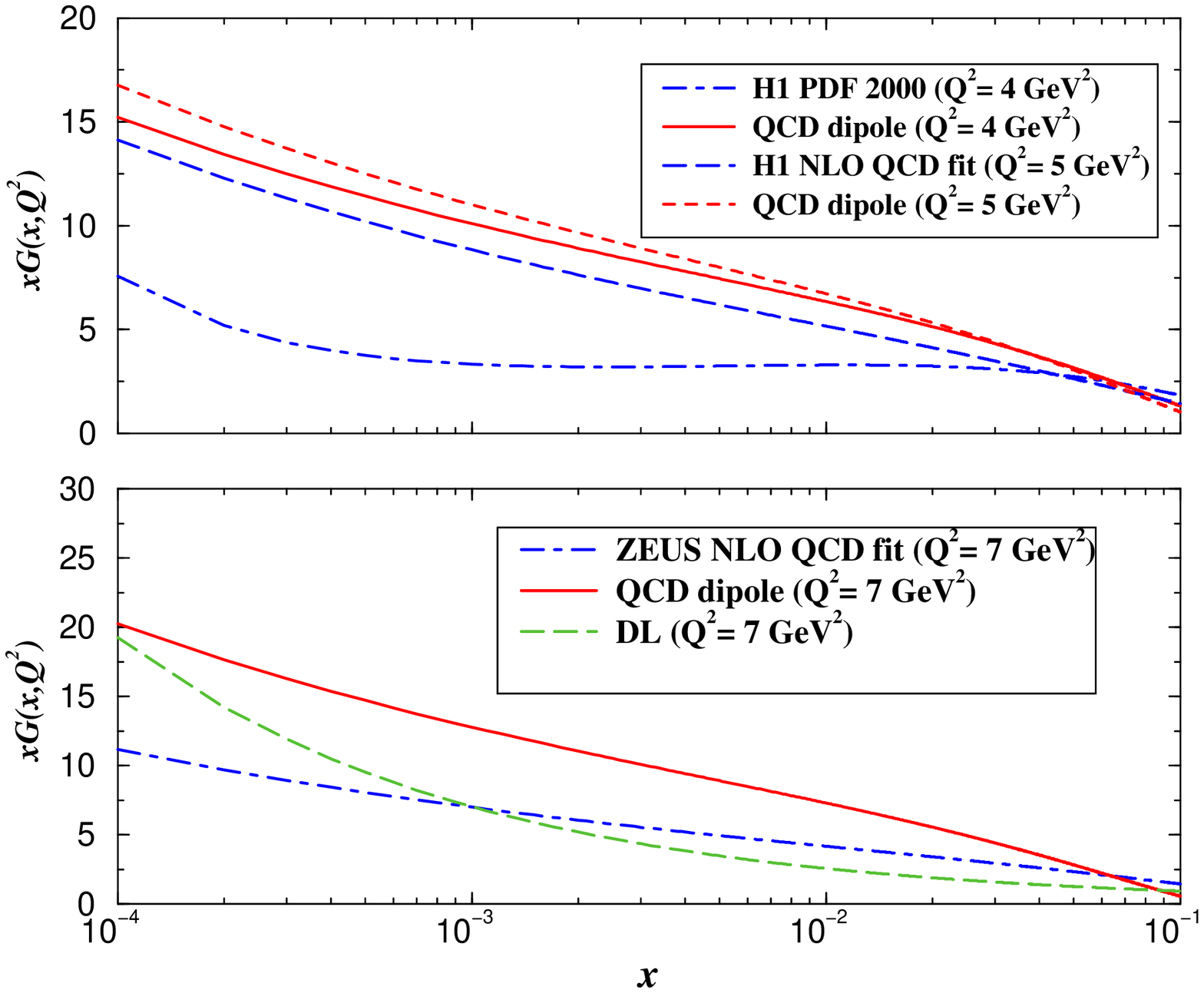,width=14cm,height=13cm}}
\caption{The gluon distribution function $x\,G(x,Q^2)$: comparison among H1 and ZEUS QCD DGLAP fit and the QCD dipole prediction.}
 \label{fig8}
\end{figure}

In Fig. (\ref{fig7}) we shown the gluon distribution function obtained using Eq. (\ref{xgxq2}) and the parameters of the procedure (III) for the H1 and ZEUS data sets. The curves represent $x\,G(x,Q^2)$ as a function of $x$ for distinct virtualities, 1.5, 15 and 150 GeV$^2$. It would be timely to compare those curves with the ones from using a resummed kernel. In order to compare these predictions with NLO QCD DGLAP fit results, in Fig. (\ref{fig8}) one plots the QCD dipole predictions with the recent QCD analysis. In the upper plot, we show the results for the H1 QCD fit (central value, without bands) in Ref. \cite{H1fitnew} for $Q^2=4$ GeV$^2$ and the earlier analysis in Ref. \cite{H1fitold}. The deviation between the recent H1 fit and the QCD dipole curves (parameters from procedure III) is sizeable, whereas it is consistent with the previous analysis. Deviations are found in the comparison with the ZEUS NLO QCD fit \cite{ZEUSfitnew} at $Q^2=7$ GeV$^2$, in the lower plot. However, it should be stressed that the gluon distribution is an indirect observable, and its determination is model-dependent. For sake of comparison with other determinations of the gluon distribution, we have plotted the result using the Regge-like hard Pomeron contribution \cite{DOLA}, which it is shown in the lower plot for $Q^2=7$ GeV$^2$. Such approach  shows that  the ratio of the gluon distribution to the hard-Pomeron part of the singlet quark distribution is about $8\pm 1$ in a large range on virtualities. Furthermore, the dependence on $Q^2$ of the hard-Pomeron contribution presents very good agreement with DGLAP evolution for $Q^2\geq 5$ GeV$^2$. 

\section{Conclusions}

We have revisited the phenomenology to the structure functions in the QCD dipole picture, performing a fit to the recent data on the proton structure function. The model for the hard Pomeron provides a reduced number of free-parameters and the results produces a good quality fit. The adjusted constants are the overall normalization to $F_2$,  the effective Pomeron intercept $\alpha_{\pom}=1+\omega_{\pom}$, the average transverse diameter of the primary dipoles in the proton $r_0=2/Q_0$ and $x_0$ scaling the $\ln (1/x)$ behavior. Moreover, the longitudinal structure function and the gluon distribution function can be determined without further adjustments. The results for $F_L$ are in good agreement with the recent measurements.  The main shortcoming from the approach is a low value for the strong coupling constant, $\alpha_s\simeq 011$, showing that a next-to-leading level analysis must be performed. The number of initial dipoles is either small, $n_{eff}=3$, and they have  a large size $r_0\simeq 1$ fm. Resummed perturbative kernels should provide a consistent value for the strong coupling constant in the range of virtualities considered, as reported in Ref. \cite{Peschanski}.  In this sense, the model presented here, although providing a remarkable data description, can be considered as effective and further phenomenology should consider higher order effects. Regarding the gluon distribution, the result is distinct from the recent NLO QCD fits, producing higher values at small $x$. We have also made a simple ansatz for describe the charm structure function, obtaining a very rough result. This is due we are disregarding the dependence of the photon impact factors on the quark mass, which modifies the final result in contrast with the assumption of avoiding the quark mass.

\section*{Acknowledgments}
M.V.T.M. thanks the support of the  High Energy Physics Phenomenology 
Group (GFPAE, IF-UFRGS) at Institute of Physics, Porto Alegre.

\end{document}